# Accurate semiempirical analytical formulas for spontaneous polarization by crystallographic parameters of SrTiO₃-BaTiO₃ system by *ab initio* calculations

Yukio Watanabe

*Department of Physics, Kyushu University, Fukuoka, Japan 819-0395*

Spontaneous polarizations ($P_S$'s) of BaTiO₃ and SrTiO₃ under various conditions are calculated *ab initio* using different exchange-correlation functionals. The extensive theoretical sets of $P_S$ vs. ion positions are found to lie on a single curve, despite the chemical differences and the wide variations of $P_S$ and lattice parameters. This uncovers accurate simple analytical formulas of $P_S$ of SrTiO₃-BaTiO₃ system expressed by ion positions; a single formula predicts both macroscopic and atomic-scale $P_S$ of SrTiO₃, BaTiO₃ and SrTiO₃-BaTiO alloys. The accuracy of the formula is demonstrated by the application to experiments, BaTiO₃-SrTiO₃ (-CaTiO₃) alloys, Sr₄Ti₄O₁₂ with $P_S$ // *a*-axis, a parallel domain, and a headon domain. In addition, the present results verify empirically that oxygen displacement is the primary identifier and the origin of $P_S$ of SrTiO₃ and BaTiO₃ and indicate that BaTiO₃ and SrTiO₃ may transforms into new state by an extremely large strain, e.g., -3%. Furthermore, the earlier prediction of headon domain without aid of defects was confirmed. The present procedures for finding formulas can be applied to other materials.

# 1. Introduction

Electrically reversible spontaneous polarization $P_S$ is the principal property of ferroelectrics [1]. $P_S$ depends on chemical composition, lattice parameters, growth processes, and ambient conditions such as temperature and strain. Although $P_S$ is estimated by polarization loop measurements or pyroelectric current [1], these measurements cannot work occasionally because of high conductance, strong pinning, or small sizes. In such cases, crystallographic estimation of $P_S$ can be useful and effective for ultrahigh speed measurements of $P_S$. Moreover, the value of $P_S$ varies within a given material due to domains, surface, interface with other materials, electric field, and strain. In these cases, the estimation of local $P_S$ or atomic scale $P_S$, is essential for the understanding ferroelectric properties. Such measurements are frequently performed through the identifications of ion positions by transmission electron microscopy (TEM) [2].

The ferroelectricity of $ABO_3$–type perovskite oxides (A: A-site ion, B: B-site ion, O: oxygen) are conventionally explained by displacement of ions from the positions of a centrosymmetric structure (Fig. 1), where the displacement patterns are described by Slater [3], Last [4] and Axe mode [5]. These displacements in $BaTiO_3$ are considered to occur in a cage formed by Ba ions [6], which suggests that the ferroelectricity can be described by the displacements of Ti and O ions. Consequently, the position of B-site ion such as Ti and Zr is used to represent local $P_S$ [7]. On the other hand, Bilz et al. [8] have successfully explained the ferroelectricity by O-ion displacements.

Indeed, we found that the plot of $P_S$ vs. the distance between Ti and O ion of $BaTiO_3$ under various strains calculated with several exchange-correlation (XC) functionals lay precisely on a single curve, despite widely varying $P_S$ [9]. The present paper reveals that all the sets of $P_S$ and certain ion positions of prototypical ferroelectrics $BaTiO_3$ and $SrTiO_3$ lie precisely on a single curve, despite chemical difference, structural difference, and widely varying $P_S$. This is probably due to the similarity between $BaTiO_3$ and $SrTiO_3$ that the bonding between A-site ion and Ti is weak [6].

The variations of $P_S$ were mostly due to some ion positions. Therefore, these plots uncover analytical expressions of $P_S$ by ion positions, e.g., given by experiments, of which validity is confirmed in Sec. 4. These examinations show also that the oxygen position is more appropriate than the Ti position as an identifier of $P_S$ consistently with Bilz et al. [8], which is more evident in $SrTiO_3$ than in $BaTiO_3$. The present paper shows also the distance between Ti and O as the primary order parameter.

## 2. Computational method

Under stress-free condition, BaTiO$_3$ is tetragonal ($C_{4v}^1$) at room temperature (RT) (Table 1 and Fig. 1(a)). RT-phase and low-temperature (low-$T$) phase SrTiO$_3$ under stress-free condition are cubic ($O_h^1$) and tetragonal ($D_{4h}^{18}$), respectively [10, 11]. The RT-phase SrTiO$_3$ can be referred to SrTiO$_3$ with the antiferro-distortive (AFD) rotation angle ($\varphi$) = 0.

As for ferroelectric distortions, small uniform displacements of ions // $c$-axis, i.e., Slater-mode-like-displacements // $c$-axis, are applied as an initial perturbation for $P_S$ except for the calculations in Sec. 4 (Table 2). Consequently, all these phases are tetragonal under biaxial inplane-stress (Table 1).

The unit-cell of low-$T$, i.e., 105K phase SrTiO$_3$ consists of 20 atoms [10] (Table 1), but the results of low-$T$ phase SrTiO$_3$ are presented in pseudo cubic unitcell [11] (Fig. 1(b)). That is, all the results are presented for 5-atom unit cell, and $P_S$ // $c$-axis except for two data in Sec. 4. The suffix "pc" stands for pseudo cubic unitcell. For example, we use $[100]_{pc}$ for the $a$-axis of pseudo cubic unitcell.

The XC functionals [12-19] used in the present calculations are listed in Table 3. Although the XC functionals used for each data point are shown in legends of Figs. 3-6, *the distinction of the XC functionals is unnecessary in the present work*. In HSE (HSE06) [18], screening length $\mu$ was the default value (0.2 Å$^{-1}$) [16, 17]. TPSS + $U$ agreed best with experiments of stress-free BaTiO$_3$ among the examined density functional +$U$ approaches [9]. Here, both Liechtenstein [19] and Dudarev [20] schemes for TPSS+$U$ without exchange term ($J$ = 0) yielded mutually similar results. Therefore, the present paper reports the results with Liechtenstein scheme with $J$ = 0 [9].

All calculations were performed using the projector augmented wave (PAW) method [21] as implemented in the Vienna *ab-initio* simulation package (Vasp) [22] with a plane wave energy cutoff of 650 eV. PAW potentials were chosen from those supplied by Vasp, so that the calculations with each XC functional agreed best with stress-free BaTiO$_3$. For the Brillouin-zone integration, Monkhorst-Pack meshes [23] of 6×6×6 and 6×6×4 were used in geometry relaxation and Berry phase calculation [24] of $P_S$ for 5 and 20 atom unit cell, respectively. In the calculations in Sec. 4, Monkhorst-Pack meshes [23] of $8\times8\times4$, $6\times6\times2$, and $2\times6\times6$ were used for BaSrTiO$_2$O$_6$, a headon-domain, and a parallel domain, respectively. In the calculations of biaxially strained cases, geometry was fully relaxed under the constraint of fixed values of the *a*-axis lattice constant. After the geometry relaxations, all the calculated forces were always less than 0.001 eV/Å, except for the calculations of domains in Sec. 4 where a few atoms in the middle of each domain were fixed. These results were the same for denser mesh and more strict criteria of calculated forces. The effects of different PAW potentials and the screening length of HSE and HSEsol were much smaller than the difference of XC functionals. The consistency and accuracy of the present calculations were reported [9, 25].

The external strain $u_{ext}$ for each phase is defined by $u_{ext} \equiv (a - a_0)/a_0$ where the theoretical *a*-axis lattice constant $a_0$ of the stress-free *minimum-free-energy* structure of each phase is calculated with *each* XC functional. The results below are for $-0.02 \leq u_{ext} \leq 0$ except for low-*T* phase SrTiO$_3$ and PBEsol calculations [9]: In the PBEsol calculations of BaTiO$_3$ and RT-phase SrTiO$_3$, $-0.04 \leq u_{ext} \leq 0$, while $-0.01 \leq u_{ext} \leq 0$ for low-*T* phase SrTiO$_3$. The $\varphi$ dependence of low-*T* phase SrTiO$_3$ was calculated by PBEsol for $0° \leq \varphi \leq 8°$ [25] in Sec. 3. The illustrations of lattices and electron densities are drawn by VESTA [26]. Figure 1(a) shows the ion positions, where O1 and O2 stands for the oxygen ions in BaO (SrO) and TiO$_2$ plane, respectively.

### 3. Crystallographic estimation of $P_S$

#### 3.1 Empirical formula of $P_S$ from ab initio results

The electronic ground state of a given ferroelectric is calculable, when the positions of all the nuclei are determined, because the ground state electron density $\rho$ is calculable from these positions (Fig.2). Therefore, $P_S$ at the ground state, which is due to the non-centrosymmetric electron and ion distribution, is calculable, when the positions of all the nuclei are known.

From now, we focus on the periodic systems, i.e., infinite single crystals. In these cases, the positions of all the nuclei *in a unitcell* are sufficient for the preceding conclusions. Therefore, "ion positions" below means "all the ion positions in a unitcell" of a given ferroelectric material. According to the preceding conclusions, when both of the electron density $\rho$ calculated for given ion positions and the calculation of polarization [24] are exact, $P_S$ should be a well-defined single-valued function of ion positions *f*: $P_S = f$(ion positions).

Indeed, we can assume that the calculations of both $\rho$ and the polarization are accurate, because $P_S$ calculated with all the XC functionals using experimental ion positions and lattice constants at 303K [27] agreed with the experimental $P_S$ at 298 K (exp.: 0.25 C/m$^2$ [28]) [9]. In addition, $P_S$ by different XC functionals was virtually the same when the crystallographic parameters were the same for all the calculations [9], although $P_S$ varied with crystallographic parameters and XC functionals (Figs. 2 and 3). Therefore, the wide variations of the *ab initio* $\rho$ and $P_S$ in Figs. 2 and 3 are considered predominantly due to the difference in the estimation of the ion positions.

The next question is "how to find the function $f$ or the most effective ion positions for $P_S$. Exemplarily in Fig. 3, each XC functional tends to yield specific deviations of crystallographic properties from ideal experimental ones, because it underestimates or overestimates some interactions depending on the approximations that each XC functional uses. Such deviations of ionic positions from ideal positions are considered to occur also locally in experiments owing to ambient temperature, defects and local strains. Therefore, the *ab initio $P_S$ deviating from ideal experiments can be reinterpreted as exact calculations of $P_S$ of the lattice deviating from ideal ones*. That is, we regard that all the *ab initio* $P_S$'s in the present calculations correspond exactly to some experimental situations or all the theoretical data in Figs. 3 - 6 represent some experiments.

In other words, the correlations between theoretical $P_S$ and theoretical crystallographic properties correspond to those in experiments. Therefore, these correlations are used to identify the primary crystallographic property responsible for $P_S$ that can be used for crystallographic identifications of $P_S$ in experiments.

Consequently, the fitting to the data in each figure yields a semiempirical analytical formula $f$ for $P_S$ expressed by crystallographic parameters. *If the selected crystallographic parameter represents the ion positions essential and necessary for $P_S$, all the data of $P_S$ vs. of this parameter should be on a single curve.* These arguments show that formulas of $P_S$ expressed by crystallographic parameters ($f$'s) exist for a given ferroelectric material.

Furthermore, in both SrTiO$_3$ and BaTiO$_3$, $P_S$ is expected to be a function of Ti, O1 and O2 positions (Fig. 1(a)), because the contribution of Ba is shown to be small [6] and that of Sr is probably also small.

Therefore, an analytical formula $f$ that works for *both* SrTiO$_3$ and BaTiO$_3$ may exist. Indeed, Fig. 6 shows that a single formula applicable for SrTiO$_3$ and BaTiO$_3$ exists, although it was unexpected.

*3.2 Overview of results in Sec. 3*

Section 3 presents the following results.

- Figure 2 illustrates how polar charge distributions appear and grow and how the difference between BaTiO$_3$ and SrTiO$_3$ originates.

- Figure 3 shows strain-dependence of $P_S$ and ion positions, which suggests the primary ions responsible for $P_S$. In addition, Fig. 3 demonstrates exemplarily wide variations of $P_S$ and ion positions due to XC functionals.

- Figure 4 examines conventionally accepted crystallographic parameters for $P_S$. These parameters are used frequently, because they are obtained by basic experiments.

- Figure 5 examines the representations of $P_S$ by the position of a single ion, where Ti position is mostly used currently.

- Figure 6 examines the positions of multiple ions for the representation of $P_S$.

In Figs. 3-6, the legends and captions explain XC functionals, e.g., PBEsol, and "_LT" in Figs. 4-6 shows a result for low-$T$ phase SrTiO$_3$. However, they can be completely ignored for the understanding of the primary crystallographic parameters, while the results for BaTiO$_3$ and those for SrTiO$_3$ should be distinguished.

*3.3 Dependence of $P_S$ on crystallographic parameters*

The electron density distribution $\rho$ in Fig. 2 shows that Ti-O distance changes with $|u_\text{ext}|$ in both BaTiO$_3$ and RT-phase SrTiO$_3$, which explains the origin of the increase of $P_S$. These displacements change chemical bonding between ions, which was evident for $|u_\text{ext}| > 0.03$ as seen in Figs. 2(c), 2(f), 2(i) and 2(l). In particular, Ti-O2, Ti-O1 and Ba (Sr)-O1 bonding (Fig. 1(a)) were evidently fortified for $|u_\text{ext}| > 0.03$. Consequently, the bonding Ba, O1 and Ti are connected by strong bonding, and the charge distribution around Ti changes from spheres to pancakes as seen in the [110] planes (Figs. 2(f) and 2(l)).

Despite the clear difference in $\rho$ at $u_\text{ext} = 0$ (Figs. 2(a), 2(d), 2(g) and 2(j)), $\rho$ of BaTiO$_3$ and RT-phase

SrTiO$_3$ are mutually similar at these large strain (Figs. 2(c), 2(f), 2(i) and 2(l)). This is consistent with $P_S$ and $\Delta z_{Ti-O2}$ of RT-phase SrTiO$_3$ similar to those of BaTiO$_3$ in Fig. 6(a). In addition, $\rho$ of BaTiO$_3$ at $u_{ext}$ = 0 resembles $\rho$ of RT-phase SrTiO$_3$ at $u_{ext}$ = -0.01 (Figs 2(a) and 2(h)), which is consistent with $P_S$ of RT-phase SrTiO$_3$ at $u_{ext}$ = -0.01 similar to $P_S$ of BaTiO$_3$ at $u_{ext}$ = 0 in Fig. 3(a)) and ref. [9] (both by PBEsol). However, the difference between BaTiO$_3$ and SrTiO$_3$ still exists at low $\rho$'s in Figs. 2(d) and 2(k).

In Fig. 3(a), the $P_S$ - $u_{ext}$ curves of RT-phase SrTiO$_3$ exhibit wide variations depending on XC functional. The displacements of Ti, O1 and O2 atom from cubic positions along *c*-axis of pseudo-cubic 5-atom unit cell in fractions of the *c* lattice constant ~ 4Å are denoted by $\Delta z_{Ti}$, $\Delta z_{O1}$, and $\Delta z_{O2}$, respectively (Table 2), and $\Delta z_{Ti-O2}$ = $\Delta z_{Ti}$ − $\Delta z_{O2}$ and $\Delta z_{Ti-O1}$ = $\Delta z_{Ti}$ − $\Delta z_{O1}$. Throughout the present paper, *we use the coordinate system in which Ba or Sr atom is always fixed at the origin (0, 0, 0) in both paraelectric and ferroelectric phase.* For example, the *z* coordinate of Ti ion is in paraelectric ($O_h^1$, P4mmm) and ferroelectric phases ($C_{4v}^1$) are 0.5 and 0.5 + $\Delta z_{Ti}$, respectively (Table 2).

In Fig. 3(b), the ionic displacements ($\Delta z_{Ti}$, $\Delta z_{O2}$, $\Delta z_{Ti-O2}$) vs. $u_{ext}$ curves depend critically on XC functionals but correspond excellently to the $P_S$- $u_{ext}$ curves of each XC functional. Similar results were found for BaTiO$_3$ [9]. This suggests that these ion displacements can be good representations for $P_S$.

In agreement with established knowledge, *c/a*, *c*-lattice constant, and $\Delta z_{Ti}$ of BaTiO$_3$ correlated with $P_S$ in ref. [9]. In RT- and low-*T* phase SrTiO$_3$, on the contrary, the correlations of $P_S$ with *c/a*, *c* and $\Delta z_{Ti}$ are ambiguous and are scattered by XC functional dependence in Figs. 4 and 5(a). Therefore, *c/a*, *c*-lattice constant, and $\Delta z_{Ti}$ are not the primary crystallographic parameter responsible for $P_S$ of SrTiO$_3$ (Nonetheless, the *c*-lattice constant in the limited range > 0.392 nm can be used to estimate $P_S$ of SrTiO$_3$ very approximately).

In Figs. 5(b) and 5(c), both oxygen ion displacements $\Delta z_{O1}$ and $\Delta z_{O2}$ correlated with $P_S$ markedly better than $\Delta z_{Ti}$. In Fig. 5(b), the $P_S$ - $\Delta z_{O1}$ plots of BaTiO$_3$ and RT- and low-*T* phase SrTiO$_3$ lie mostly on a single curve. Especially, the plots of $P_S$ - $\Delta z_{O1}$ of RT- and low-*T* phase SrTiO$_3$ lie on a straight line. In Fig. 5(c), the plots of $P_S$ - $\Delta z_{O2}$ of BaTiO$_3$ lie on a single curve, while the plots of RT- and low-*T* phase SrTiO$_3$ lie excellently on another single line. This indicates that $\Delta z_{O2}$ is an essential part of the primary crystallographic property for $P_S$ and can be used for the estimation of $P_S$.

In Fig. 6(a), all the $P_S$ - $\Delta z_{Ti-O2}$ plots lie on a single straight line, despite the difference in XC functionals, the magnitude of strain, AFD rotation angle $\varphi$, lattice, symmetry, and material. In addition, all the $P_S$ - $\Delta z_{Ti-O1}$ plots lie almost on a single curve in Fig. 6(b). Therefore, $\Delta z_{Ti-O2}$ and $\Delta z_{Ti-O1}$ are primary crystallographic properties for $P_S$. These correlations yield analytical formulas of $P_S$ expressed by $\Delta z_{Ti-O2}$ or $\Delta z_{Ti-O1}$, which is summarized in Table 4. Almost the same results as Figs. 5 and 6 were obtained changing the unit of the displacements, e.g., $\Delta z_{Ti-O2}$ from fraction of $c$-lattice constant to absolute value (Å).

As mentioned above, the ground state electron density $\rho$ is determined from the ion positions. Therefore, even when we claim that $P_S$ is determined by the displacement of some ions, the origin of $P_S$ is still ion and electrons. Under this presumption, the ferroelectricity of BaTiO$_3$ and SrTiO$_3$ is considered as oxygen-driven, because $P_S$ correlated excellently with $\Delta z_{O2}$ and $\Delta z_{Ti-O2}$ and the main contribution to $\Delta z_{Ti-O2}$ is from $\Delta z_{O2}$. This is mainly due to the value of $|\Delta z_{O2}|$ and $|\Delta z_{O1}|$ far larger than $\Delta z_{Ti}$ as seen in Fig. 5.

The oxygen mechanism is considered more dominant in SrTiO$_3$ than in BaTiO$_3$, because $P_S$ correlated with $\Delta z_{O2}$ and $\Delta z_{O1}$ better in SrTiO$_3$ than BaTiO$_3$ and correlated with $\Delta z_{Ti}$ worse in SrTiO$_3$ than BaTiO$_3$ (Figs. 5(b) and 5(c)). In terms of $\Gamma 15$ zone-center phonon of cubic lattice, the present displacement patterns were equal to a mixture of Slater, Last and Axe modes [3-5], whereas the weight of Slater mode was larger in BaTiO$_3$ than the others, and the weight of each mode was nearly the same in the ferroelectricity of SrTiO$_3$.

In Figs. 5 and 6, the results for SrTiO$_3$ and BaTiO$_3$ at $u_{ext}$ = -0.03 and -0.04, which are shown by small filled red circles and small half-filled brown circles, respectively, exhibit the $u_{ext}$-dependence different from those for $|u_{ext}| \leq 0.02$. Here, the growth of SrTiO$_3$ and BaTiO$_3$ at $u_{ext} \leq$ -0.03 without defects is experimentally difficult. Indeed, Fig. 3(b) suggests that the strain dependence of $\Delta z_{Ti}$, $\Delta z_{O2}$ and, $\Delta z_{Ti-O2}$ changes for $u_{ext} \leq$ -0.03; for example, $\Delta z_{Ti}$ decreased with $|u_{ext}| \geq 0.03$ in Fig. 3(b). In addition, the unit cell volume increased with $|u_{ext}| \geq 0.03$, while it decreased with $|u_{ext}| \leq 0.02$. These characteristics were also found in BaTiO$_3$ [9]. Therefore, SrTiO$_3$ and BaTiO$_3$ for $u_{ext} \leq$ -0.03 are not included in the fitting.

### 4. Validity of the present formulas

Among the formulas obtained in Sec. 3, the formulas based on $\Delta z_{Ti-O2}$ and $\Delta z_{Ti-O1}$ were accurate, while the formula based on $\Delta z_{O2}$ was also accurate when different formulas are used for SrTiO$_3$ and BaTiO$_3$. Because the formulas based on $\Delta z_{Ti-O2}$ and $\Delta z_{Ti-O1}$ work for both SrTiO$_3$ and BaTiO$_3$, the validity of the following formulas (Table 4) are examined with experimental values and *ab initio* calculations:

$$P_S \,(\mu C/cm^2) = 977 \Delta z_{Ti-O2} + 0.04, \qquad (1)$$

$$P_S \,(\mu C/cm^2) = 867 \Delta z_{Ti-O1} - 2730 \Delta^2 z_{Ti-O1} + 0.82 \qquad (2)$$

Because Eqs. (1) and (2) use $\Delta z_{Ti-O2}$ and $\Delta z_{Ti-O1}$, respectively, we define an oxygen position using "O*": O* in $\Delta z_{Ti-O2}$ (O*) is an O on a Ti-O bonding perpendicular to the direction of a $P_S$ component, and O* in $\Delta z_{Ti-O1}$ is an O on a Ti-O bonding along a $P_S$ component. For example, O* of $\Delta z_{Ti-O2}$ is O2 of Fig. 1(a) for $P_S$ // [001]$_{pc}$ and O1 of Fig. 1(a) for $P_S$ // [110]$_{pc}$.

When the unitcell consists of multiple pseudo cubic cells as in BaSrTi$_2$O$_6$ and low-*T* phase SrTiO$_3$ in Fig. 7, *$P_S$ is defined as an average of all the pseudo cubic cells*. When $P_S$ // [001]$_{pc}$ ("pc" stands for pseudo cubic),

$\Delta z_{Ti-O2}$ = distance between Ti and O* ion (Å) ÷ *c*-lattice constant of pseudo cubic 5-atom cell (Å).

For simplicity, the *c*-lattice constant of pseudo cubic cells can be approximated by an average of all the pseudo cubic cells: For example, the *c*-lattice constant of pseudo cubic cells in BaSrTi$_2$O$_6$ is a half of the *c*-lattice constant of BaSrTi$_2$O$_6$.

Using these definitions, we examine the $P_S$ obtained by Eqs. (1) and (2). The first comparison is that with experiments: In Fig. 7, the two black squares represent the comparison between experimental $P_S$ of BaTiO$_3$ [28] and $P_S$ estimated by Eqs. (1) and (2) using experimental $\Delta z_{Ti-O2}$ and $\Delta z_{Ti-O1}$ of BaTiO$_3$ [27].

The second comparison is the validity for the alloys of BaTiO$_3$ and SrTiO$_3$ (CaTiO$_3$): The two red circles represent the comparison of *ab initio* $P_S$ of BaSrTi$_2$O$_6$ with $P_S$ estimated by Eqs. (1) and (2) using *ab initio* $\Delta z_{Ti-O2}$ and $\Delta z_{Ti-O1}$ of BaSrTi$_2$O$_6$. Similarly, the comparisons of Ba$_2$SrTi$_3$O$_9$ and BaCaSrTi$_3$O$_9$ are shown.

The third comparison in Fig. 7 is the validity for $P_S$ // *a*-axis in the unit cell of Sr$_4$Ti$_4$O$_{12}$, where O2's on the diagonal positions (illustration of Fig. 7) are rotating around the Ti-O1 axis, i.e., *c*-axis. In this case, we assumed O* as O1 of Fig. 1(a), i.e. oxygens above and below Ti in the green-framed illustration of Fig. 7. The green diamond in Fig. 7 represents the comparison between *ab initio* $P_S$ and $P_S$ estimated by Eqs. (1) and

(2) using *ab initio* $\Delta z_{\text{Ti-O2}}$. All the three comparisons confirm the validity of Eqs. (1) and (2).

The variation of local $P_S$ within parallel and headon domains of BaTiO$_3$ is shown in Figs. 8(a) and 8(b), where the local $P_S$'s were estimated by Eqs. (1) and (2) and *ab initio* calculations of extracted 5-atom unitcells like Fig. 1(a). The distribution of local $P_S$ in the parallel domains agreed with literature [29, 30]. $\Delta z_{\text{Ti-O2}}$ and $\Delta z_{\text{Ti-O1}}$ used in Eqs. (1) and (2) was those of the extracted 5-atom unitcells. The *z*-axis is along *c*-axis, and the parallel domain runs along *x*-axis (*a*-axis). In both types of domains, local $P_S$'s by Eqs. (1) and (2) agreed excellently with local $P_S$'s by *ab initio* calculation. In addition, small *x*-components of $P_S$ ($P_{Sx}$) existed in parallel domain, and local $P_{Sx}$'s by *ab initio* calculation and Eqs. (1) and (2) agreed mutually. Headon domains like Fig. 8(b) were conventionally unstable but predicted to exist even in the absence of defects owing to the formation of free electron/hole layers [31, 32], of which experiments are recently frequently reported.

## 5. Summary

The data sets of $P_S$ vs. crystallographic parameters of BaTiO$_3$ and SrTiO$_3$ under various conditions calculated with different XC functionals were found to lie precisely on a single curve (Figs. 5(b), 5(c) and 6), despite the wide variations of these data (Fig. 3). Here, the origin of these variations of $P_S$ was found due to the variations of crystallographic parameters. Because such variations of crystallographic parameters were expected to exist in experiments, the parameters' theoretical variations were regarded as the experimental ones. These considerations have yielded empirical analytical formulas of $P_S$ by $\Delta z_{\text{Ti-O2}}$, $\Delta z_{\text{Ti-O1}}$, $\Delta z_{\text{O2}}$, and $\Delta z_{\text{O1}}$ (Table 4), where $\Delta z_{\text{O1}}$ and so forth are explained in Fig. 1(a) and Table 2.

The parameters that correlated best with $P_S$ were $\Delta z_{\text{Ti-O2}}$, $\Delta z_{\text{Ti-O1}}$, and $\Delta z_{\text{O2}}$, suggesting the importance of oxygen position and, hence, requesting accurate detection of oxygen positions. The coefficients in the formulas using these displacements are in Table 4.

The validity of the formulas based on $\Delta z_{\text{Ti-O2}}$ and $\Delta z_{\text{Ti-O1}}$ (Eqs. (1) and (2)) was successfully shown in comparison with experiments at RT, BaTiO$_3$-SrTiO$_3$ (-CaTiO$_3$) alloys, a complicated unit cell (Sr$_4$Ti$_4$O$_{12}$), a parallel domain, and a headon domain. As shown in Fig. 7, the present formulas such as Eq. (1) work for $P_S$ at any temperature as long as ion positions such as at that temperature is used. Therefore, ion positions given by various kinds of experiments can be used for the present formulas.

The validity of Eqs. (1) and (2) in application to BaTiO$_3$-SrTiO$_3$ alloys shows that Eqs. (1) and (2) are applicable to BaTiO$_3$-SrTiO$_3$ systems with good accuracy. However, it did not work for PbTiO$_3$, where the major difference between PbTiO$_3$ and BaTiO$_3$-SrTiO$_3$ is the bonding of A-site atom with Ti [6]. Therefore, Eqs. (1) and (2) may hold for titanate perovskites in which the bonding between Ti and A-site atom is weak, although the inclusion of Ca may decrease the accuracy of Eqs. (1) and (2).

For an extremely large strain, e.g., $u_{ext} \leq -0.03$ in both BaTiO$_3$ and SrTiO$_3$, the correlations between $P_S$ and displacements and those between properties and strain changed slightly. In addition, the unitcell volume increased with $|u_{ext}|$ for $u_{ext} \leq -0.03$, while it decreased with $|u_{ext}|$ for $u_{ext} \geq -0.02$. This suggests that BaTiO$_3$ and SrTiO$_3$ transforms into new states by an extremely large strain. This view is supported by the sharp enhancement of the bonding between atoms in both BaTiO$_3$ and SrTiO$_3$ in the charge density $\rho$ for $u_{ext} \leq -0.03$ (Fig. 2).

The present correlations of $P_S$ with $\Delta z_{O2}$ and $\Delta z_{O1}$ far better than those with $\Delta z_{Ti}$ verifies empirically oxygen-driven ferroelectricity by Bilz et al. [8] for BaTiO$_3$-SrTiO$_3$ system, where this mechanism is considered more dominant in SrTiO$_3$ than in BaTiO$_3$. Accordingly, the conventional identifier $\Delta z_{Ti}$ [7] was inaccurate and had possibility to estimate $P_S$ incorrectly. This may be important for delicate cases such as flux-closure domain at strained interface [7]. In addition, the earlier prediction of stable headon domains without aid of defects [31, 32] was confirmed.

**Acknowledgement**

The author acknowledges useful discussions with Dr. Masao Arai and Prof. P. Blöchl.

**Data availability**

The raw data required to reproduce these findings cannot be shared at this time as the data also form part of an ongoing study. The processed data required to reproduce these findings cannot be shared at this time as the data also forms part of an ongoing study.

**Appendix: How to use the formulas**

The formulas in Table 4 are for the case of $P_S \geq 0$. For $P_S < 0$, the sign of A and C should be reversed. Here, the convention of the sign of $P_S$ is such that $P_S > 0$ for Ti –O2 > 0, which may be opposite to the convention of *ab initio* calculations.

These formulas are for $P_S$ in 5-atom unit cell, and the length of the lattice along $P_S$ is unity. In application to ferroelectrics consisting of many unitcells, local $P_S$, i.e. $P_S$ in each 5-atom unit cell is calculated, and the average of local $P_S$ gives the bulk $P_S$. For example, local $P_S$ in BaSrTi$_2$O$_6$ in Fig. 7 is calculated as follows:

By denoting the $z$ coordinates of Ba, Sr, two Ti's, two O2's and two O1's as zBa, zSr, zTi$_1$, zTi$_2$, zO2$_1$, zO2$_2$, zO1$_{Ba}$ and zO1$_{Sr}$, respectively,

$\Delta z_{Ti-O2}$ = (zTi$_1$ − zO2$_1$)/(zSr − zBa)    and

$\Delta z_{Ti-O2}$ = (zTi$_2$ − zO2$_2$)/(zBa* − zSr),

where zBa*= 1 and zBa = 0. Here, Ti-O2 distance is normalized by the local $c$ lattice constants zSr − zBa or zBa* − zSr. Alternatively, we may *approximate* (zSr − zBa) ≈ (zBa* − zSr) ≈ $c^{BST}$/2, where $c^{BST}$ is the $c$ lattice constant of BaSrTi$_2$O$_6$. Similarly, by noting that the distance between Ti and O1 in paraelectric phase is 0.5,

$\Delta z_{Ti-O1}$ = (zTi$_1$ − zO1$_{Ba}$)/(zSr − zBa) − 0.5 and

$\Delta z_{Ti-O1}$ = (zTi$_2$ − zO1$_{Sr}$)/(zBa*− zSr) − 0.5.

Local $P_S$ is obtained by the substitutions of these $\Delta z_{Ti-O2}$ and $\Delta z_{Ti-O1}$ into Eqs. (1) and (2). For general case of $P_S$ // layer-direction, e.g., headon domain (Fig. 8(b)),

$\Delta z_{Ti-O2}^k$ = (zTi$^k$ − zO2$^k$)/(zA$^{k+1}$− zA$^k$) and

$\Delta z_{Ti-O1}^k$ = (zTi$^k$ − zO1$^k$)/(zA$^{k+1}$− zA$^k$) − 0.5,

where $zA^k$ is the z coordinate of A-site ion in the $k^{th}$ unit cell. For general case of $P_S \perp$ layer-direction, e.g., parallel domain (Fig. 8(a))

$\Delta z_{Ti-O2}^k$ = (zTi$^k$ − zO2$^k$) and

$\Delta z_{Ti-O1}^k$ = (zTi$^k$ − zO1$^k$) − 0.5.

Alternatively, we may *approximate* zA$^{k+1}$− zA$^k$ ≈ $c^{TOT}$/$n$ except for the surfaces, where $c^{TOT}$ is the total $c$ lattice constant of the system, i.e., supercell and $n$ is the number of 5-atom unitcells in the supercell.

In applications of these formulas, especially to materials that are not BaTiO$_3$-SrTiO$_3$, the user should test them by comparing a few $P_S$'s given by the formulas with those given by the other methods such as experiments or Berry phase calculations.

**Table Captions**

**Table 1.**

Symmetry of $SrTiO_3$ and $BaTiO_3$ studied in the present work. The calculated $SrTiO_3$ phases are cubic ($O_h^1$), tetragonal having $P_S \mathbin{/\mkern-2mu/} c$ ($C_{4v}^1$, $D_{4h}^1$, $D_{4h}^{18}$, and $C_{4v}^{10}$), and orthorhombic ($C_{2v}^{22}$). The calculated $BaTiO_3$ phases are tetragonal having $P_S \mathbin{/\mkern-2mu/} c$ ($C_{4v}^1$). In the experiments at RT, $SrTiO_3$ is $O_h^1$ under stress-free condition and is $D_{4h}^1$ or $C_{4v}^1$ under biaxial strain. The low-$T$ $SrTiO_3$ is $D_{4h}^{18}$ in stress-free experiments. The low-$T$ $SrTiO_3$ with $P_S \mathbin{/\mkern-2mu/} c$ and $P_S \mathbin{/\mkern-2mu/} a_t$ are $C_{4v}^{10}$ and $C_{2v}^{22}$, respectively, where $a_t$ is an $a$-axis of 20-atom tetragonal cell.

**Table 2.**

Atom positions in unitcells of paraelectric cubic ($O_h^1$) $SrTiO_3$ and ferroelectric tetragonal ($C_{4v}^1$ (P4mm)) $BaTiO_3$ and $SrTiO_3$ that have $P_S \mathbin{/\mkern-2mu/} c$. In Figs. 2-6, $P_S \mathbin{/\mkern-2mu/} c$. Ba or Sr atom is always fixed at (0, 0, 0) in both paraelectric and ferroelectric phase. The unit is $a = c = 1$, where $a$ and $c$ are the $a$ and $c$ lattice constant of the unitcell, respectively. The displacements $\Delta z_{Ti}$, $\Delta z_{O1}$ and $\Delta z_{O2}$ shown by the arrows in Fig. 1(a) are measured from cubic position of each ion along $c$-axes ($z$-axis) in fractions of the $c$ lattice constant. O1 and O2 are defined in Fig. 1(a). In pseudo cubic unitcell representations of low-$T$ phase $SrTiO_3$, the $z$-coordinates of atoms of paraelectric ($D_{4h}^{18}$) and ferroelectric ($C_{4v}^{10}$ (I4cm)) that has $P_S \mathbin{/\mkern-2mu/} c$ are similar to those of the paraelectric and ferroelectric in Table 2, respectively.

**Table 3.**

List of exchange correlation (XC) functionals used in this work. However, it is unnecessary to know which XC functional was used in Figs. 2-8. "_LT" shows the calculations of low-$T$ phase $SrTiO_3$, whereas those without "_LT" are for the RT-phase.

**Table 4.**

Coefficients of fittings to *ab initio* results. $R^2$ and $\sigma$ stand for the coefficient of determination and the standard deviation, respectively. These formulas are for the case of $P_S \geq 0$. "How to use the formulas" is explained in Appendix.

**Figure captions**

**Fig. 1.** (a) Unit cell consisting of 5 atoms, while green spheres, light blue spheres, and red spheres represent Ba(Sr), Ti and O ions, respectively. The coordinates are defined such that $x$, $y$ and $z$-axis are along the $a$-, $a$- and $c$-axis of the unitcell, respectively. (b) Unit cell consisting of 20 atoms of low-$T$ phase of SrTiO$_3$ viewed along $c$-axis (top view), where the orange frame shows pseudo cubic unit cell of 5 atoms.

**Fig. 2.** Calculated valence electron density $\rho$ of ground state of (a)-(f) tetragonal BaTiO$_3$ and (g)-(l) RT-phase SrTiO$_3$ with $P_S > 0$ under strain-free and strained condition, where the value of external strain $u_{ext}$ is shown on the left. (a)-(c) [100] and (d)-(f) [110] planes of BaTiO$_3$. (g)-(i) [100] and (j)-(l) [110] plane of SrTiO$_3$. The first row: (a), (d), (g) and (j) is for $u_{ext} = 0$, the second row: (b), (e), (h) and (k) is for $u_{ext} = -0.01$, and the third row (c), (f), (i) and (l) is for $u_{ext} = -0.04$. The calculation for cubic SrTiO$_3$ is shown in (g) and (j), but the calculations with some XC functionals show the existence of $P_S$ for $u_{ext} = 0$ [25]. The red color represents $\rho > 0.5$ $e$/Å$^3$ in the [100] planes ((a)-(c) and (g)-(i)) and $\rho > 0.1$ $e$/Å$^3$ in the [110] planes ((d)-(f) and (j)-(l)), and $\rho$ decreases in the order of red, yellow, green, and blue. The positions of ions are shown by letters, where O1 and O2 are oxygen in BaO (SrO) and TiO2 layer, respectively (Fig. 1(a)).

**Fig. 3.** (a) External biaxial inplane strain ($u_{ext}$) dependence of $P_S$ of RT-phase SrTiO$_3$. Both large and small filled red circles, filled green diamonds, half-filled green diamonds, and light blue squares represent the results of PBEsol, HSE, HSEsol, and LDA for RT-phase SrTiO$_3$, respectively. This convention of symbols is kept throughout the present paper. (b) Displacements of Ti and O2 atom from cubic positions along $c$-axis ($z$-axis) ($\Delta z_{Ti}$, $\Delta z_{O2}$) and difference of them ($\Delta z_{Ti-O2}$) in fractions of the $c$ lattice constant (Table 2) are plotted against $u_{ext}$. The small symbols represent $\Delta z_{Ti}$.

**Fig. 4.** (a) $P_S$ vs. $c/a$ and (b) $P_S$ vs. $c$ lattice constant for stress-free and strained RT- and low-$T$ phase SrTiO$_3$. Solid lines are fittings to the results given by HSE. The meanings of other symbols are same as in Fig. 3. In addition, filled pink circles and filled dark blue diamonds represent the results of PBEsol and HSE results for low-$T$ phase SrTiO$_3$, respectively, where "_LT" in legends stands for "low-$T$"(Table 3).

**Fig. 5.** Correlation of $P_S$ with atomic displacements (a) $\Delta z_{Ti}$ of stress-free and strained RT- and low-$T$

phase SrTiO$_3$. Correlation of $P_S$ with (b) $\Delta z_{O1}$ and (c) $\Delta z_{O2}$ of stress-free and strained BaTiO$_3$ and RT- and low-$T$ phase SrTiO$_3$. $\Delta z_{Ti}$, $\Delta z_{O1}$ and $\Delta z_{O2}$ are defined in Table 2 and shown by the arrows in of Fig. 1(a). The distinction of XC functionals shown by symbols in the legends is unnecessary: Filled red circles, filled green diamonds, half-filled green diamonds, and light blue squares represent the results of PBEsol, HSE, HSEsol, and LDA for RT-phase SrTiO$_3$, respectively. Filled pink circles and filled dark blue diamonds represent the results for low-$T$ phase SrTiO$_3$. Half-filled brown circles, half-filled dark green diamonds, small half-filled dark green diamonds, small filled violet squares, filled blue triangle, and filled inverted blue triangle represent the results of PBEsol, HSEsol, HSE, LDA, TPSS, and TPSS+$U$ for BaTiO$_3$, respectively. In Figs. 5 and 6, the results at $u_{ext}$ = -0.03 and -0.04 (PBEsol) are shown by small filled red circles and small half-filled brown circles, which were not used in the fittings. In (b), solid orange lines are fitting (Table 4). In (c), solid dark yellow and orange lines are 2$^{nd}$ and 3$^{rd}$ order polynomial fittings to all the BaTiO$_3$ data, respectively, while solid violet lines are fitting to SrTiO$_3$ data.

**Fig. 6.** Correlation of $P_S$ with (a) $\Delta z_{Ti-O2}$ and (b) $\Delta z_{Ti-O1}$ of stress-free and strained BaTiO$_3$ and RT- and low-$T$ phase SrTiO$_3$. $\Delta z_{Ti-O2}$ and $\Delta z_{Ti-O1}$ are the difference between $\Delta z_{Ti}$ and $\Delta z_{O2}$ and that between $\Delta z_{Ti}$ and $\Delta z_{O1}$, respectively. The meanings of the symbols are the same as those in Fig. 5, but the distinction of XC functionals is unnecessary. In (a), the solid orange and dashed black lines are fitting to all the data and BaTiO$_3$ data, respectively (Table 4). In (b), solid dark yellow and orange lines are 2$^{nd}$ and 3$^{rd}$ order polynomial fittings to all data, respectively, while dashed black lines are fitting to BaTiO$_3$ data.

**Fig. 7.** Comparison of macroscopic $P_S$ estimated by Eqs. (1) and (2) with experimental and *ab initio* $P_S$ of bulk materials: BaTiO$_3$ (experiments of $P_S$ at 298 K [27] and $\Delta z_{Ti-O2}$ and $\Delta z_{Ti-O1}$ at 303 K [28]), black squares), BaSrTi$_2$O$_6$ (*ab initio*, red circles), Ba$_2$SrTi$_3$O$_9$ (*ab initio*, pink hexagons), BaCaSrTi$_3$O$_9$ (*ab initio*, light blue star), low-$T$ phase tetragonal SrTiO$_3$ with $P_S$ along [100] (*ab initio*, green diamond). The filled and open symbols represent the results by Eqs. (1) and (2), respectively. The solid lines show the curve for "$P_S$ by experiments or *ab initio* calculation = $P_S$ by the formula"; all the symbols should be on this line, when the estimations by the formula are perfectly accurate. The illustrations in red, pink, light blue and green frame show side views of BaSrTi$_2$O$_6$, Ba$_2$SrTi$_3$O$_9$ and BaCaSrTi$_3$O$_9$, and a top view of SrTiO$_3$ unit cell consisting of 25 atoms, respectively. In each illustration, green spheres, dark green spheres, blueish

green spheres, light blue spheres, and red spheres represent Ba, Sr, Ca, Ti and O ions, respectively, and the black arrow shows the direction of $P_S$.

**Fig. 8.** Estimations of atomic-scale, i.e., local $P_S$. (a) parallel, i.e. *cc* domain running in the direction of *a*-axis (*x*), where the entire upward and downward domain consist of 16 unitcell (80 atoms). Filled black squares, filled red circles, open pink circles represent $P_S$ component along *c*-axis (*z*) estimated by *ab initio* calculation, Eq. (1), and Eq. (2), respectively. Filled green squares, filled blue circles, and open light blue circles represent $P_S$ component along *a*-axis estimated by *ab initio* calculation, Eq. (1), and Eq. (2), respectively. (b) Headon, i.e. encountering domain running in the direction of *c*-axis (*z*), where the entire upward and downward domain consist of 20 unitcell (100 atoms). Filled black squares, filled blue circles, and open green circles represent $P_S$ estimated by *ab initio* calculation, Eq. (1), and Eq. (2), respectively.

Table 1

| SrTiO3 | paraelectric | ferroelectric $P_S$ // $c$ | ferroelectric $P_S$ // $a$ |
|---|---|---|---|
| RT-phase | $O_h^1$ (Pm3m) ($D_{4h}^1$ (P4mmm)) | $C_{4v}^1$ (P4mm) | |
| low-$T$-phase | $D_{4h}^{18}$ (I4mcm) | $C_{4v}^{10}$ (I4cm) | $C_{2v}^{22}$ (Ima2m) |
| RT-BaTiO3 | | $C_{4v}^1$ (P4mm) | |

Table 2

| | Paraelectric ($O_h^1$, $D_{4h}^1$) | Ferroelectric ($C_{4v}^1$) |
|---|---|---|
| Ba, Sr | (0, 0, 0) | (0, 0, 0) |
| Ti | (0.5, 0.5, 0.5) | (0.5, 0.5, 0.5 + $\Delta z_{Ti}$) |
| O1 | (0.5, 0.5, 0) | (0.5, 0.5, 0 + $\Delta z_{O1}$) |
| O2 | (0, 0.5, 0.5) | (0, 0.5, 0.5 + $\Delta z_{O2}$) |
| O2 | (0.5, 0, 0.5) | (0, 0.5, 0.5 + $\Delta z_{O2}$) |

Table 3

| name | description |
|---|---|
| LDA | Local density approximation |
| PBEsol | Perdew-Burke-Ernzerhof functional for solids [13] |
| TPSS | Tao-Perdew-Staroverov-Scuseria functional [14,15] |
| HSE | Heyd-Scuseria-Ernzerhof functional (HSE06) [16,18] |
| HSEsol | HSE for solid [17] |
| TPSS + $U$ | TPSS with Hubbard-like local potential [19] |

| PBEsol_LT | PBEsol calculation of low-$T$ SrTiO3 : Figs.4-6 |
|---|---|
| HSE_LT | HSE calculation of low-$T$ SrTiO3 : Figs.4-6 |

Table 4

| $P_S(\mu C/cm^2) = A+Bx+Cx^2+Dx^3$ | | $x$ | A | B | C | D | $R^2$ | $\sigma$ |
|---|---|---|---|---|---|---|---|---|
| Fig.5(b) | All | $\Delta z_{O1}$ | -0.77 | -1670 | -22600 | -106000 | 0.982 | 0.0025 |
| Fig.5(c) | BTO | $\Delta z_{O2}$ | 0.01 | -2280 | -22700 | 0 | 0.999 | 0.0050 |
| | BTO | $\Delta z_{O2}$ | -0.01 | -2500 | -39700 | -313000 | 0.999 | 0.0070 |
| | STO | $\Delta z_{O2}$ | 0.29 | -1040 | 0 | 0 | 0.997 | 0.0007 |
| | STO | $\Delta z_{O2}$ | 0.05 | -1130 | -3230 | 0 | 0.996 | 0.0007 |
| Fig.6(a) | BTO | $\Delta z_{Ti-O2}$ | 0.04 | 977 | 0 | 0 | 1.000 | 0.0045 |
| | STO | $\Delta z_{Ti-O2}$ | -0.05 | 926 | 0 | 0 | 0.999 | 0.0004 |
| | All | $\Delta z_{Ti-O2}$ | -0.02 | 959 | 0 | 0 | 0.999 | 0.0098 |
| Fig.6(b) | BTO | $\Delta z_{Ti-O1}$ | 8.52 | 549 | 0 | 0 | 0.988 | 0.0101 |
| | BTO | $\Delta z_{Ti-O1}$ | 0.01 | 851 | -2560 | 0 | 0.999 | 0.0063 |
| | STO | $\Delta z_{Ti-O1}$ | 0.27 | 954 | 0 | 0 | 0.995 | 0.0010 |
| | All | $\Delta z_{Ti-O1}$ | 0.82 | 867 | -2730 | 0 | 0.989 | 0.0185 |
| | All | $\Delta z_{Ti-O1}$ | -0.23 | 1190 | -13100 | 81700 | 0.994 | 0.0129 |

Fig.1

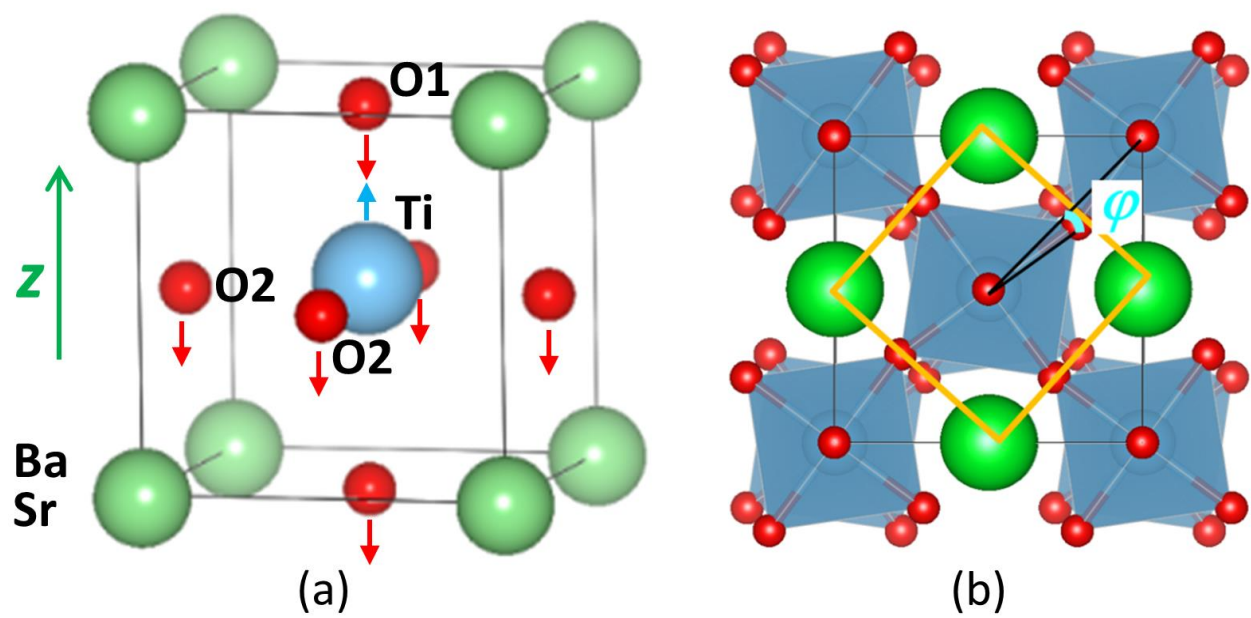

(a)　　　　　　　　　　　　(b)

**Fig.2**

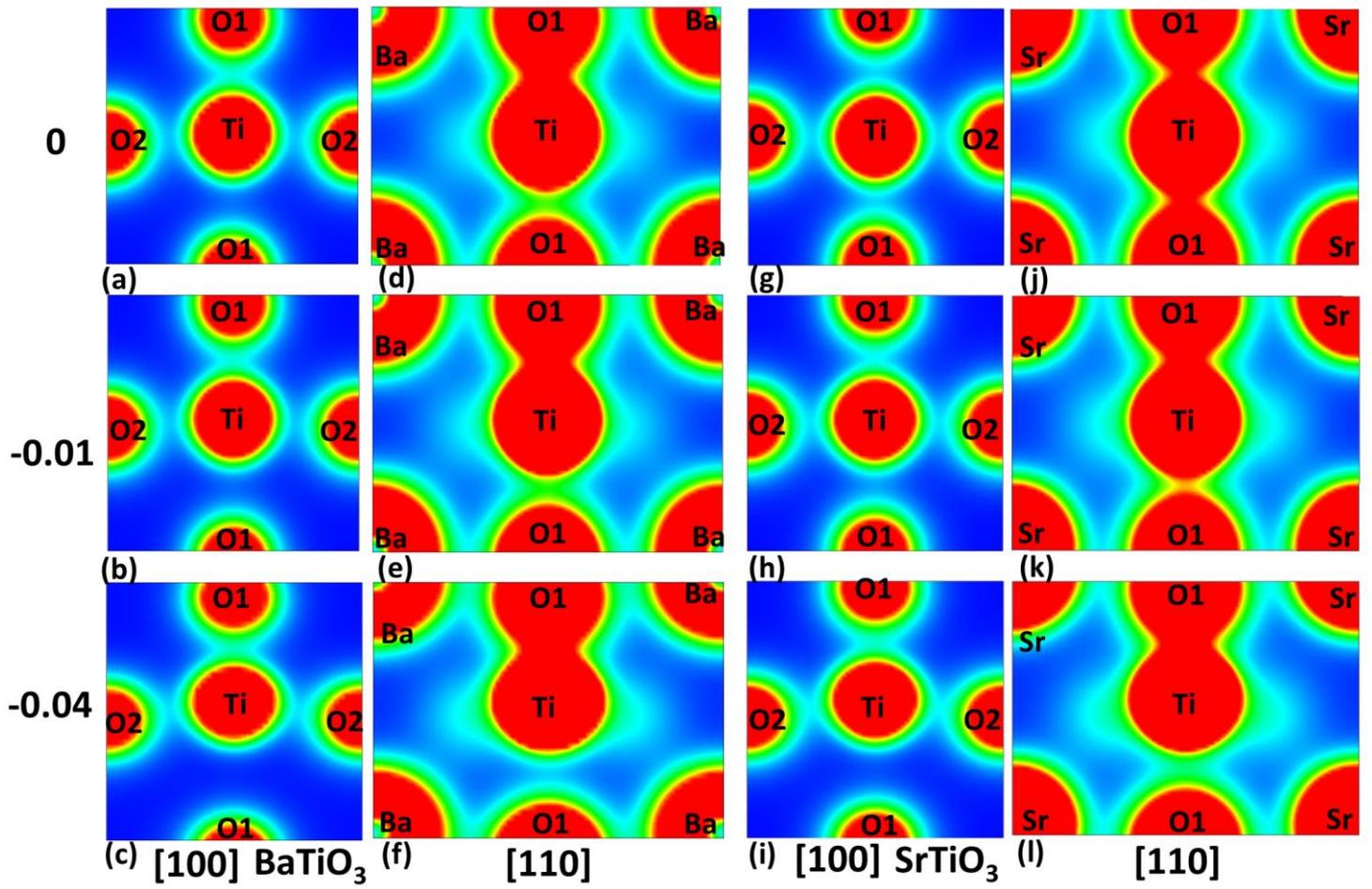

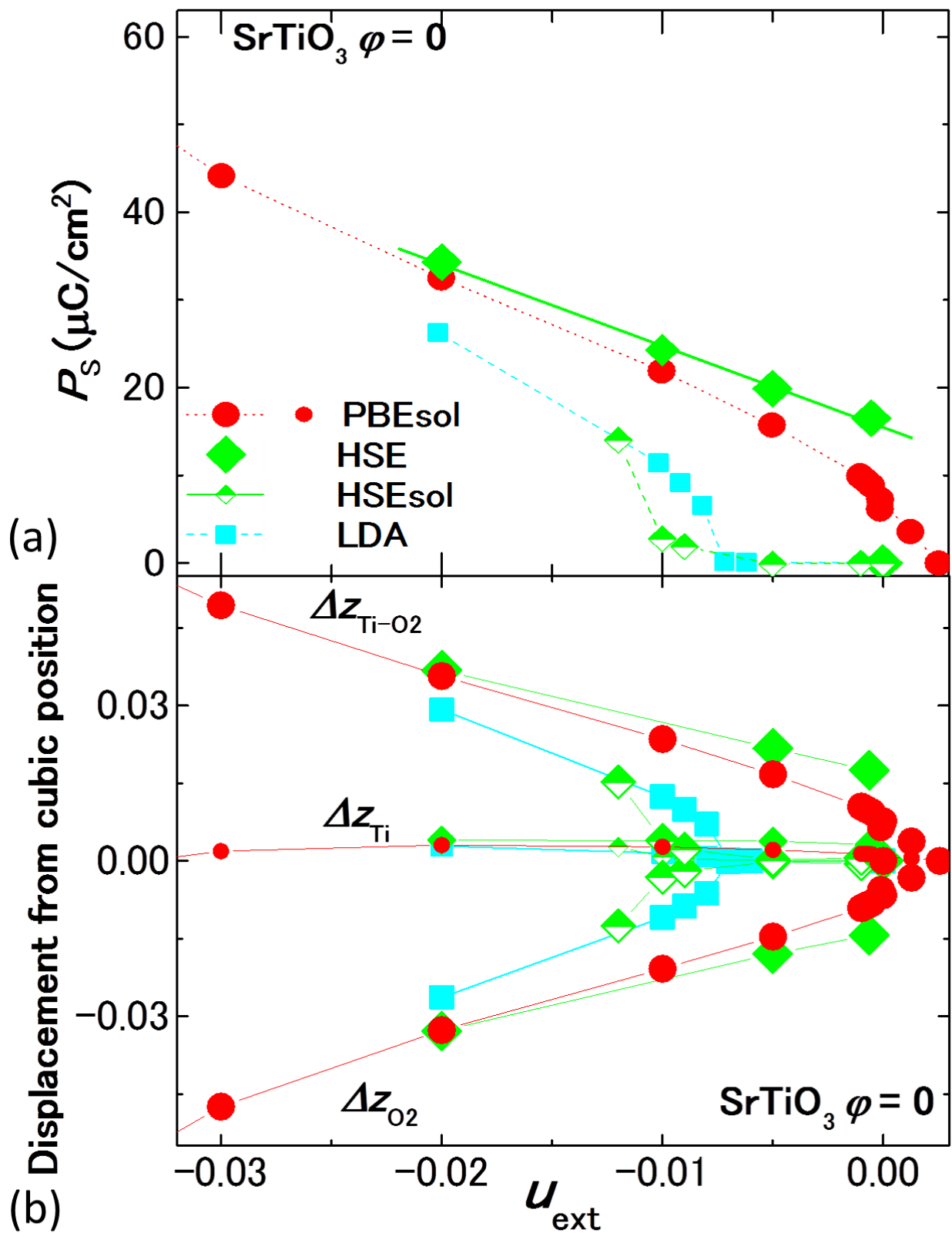

Fig.4

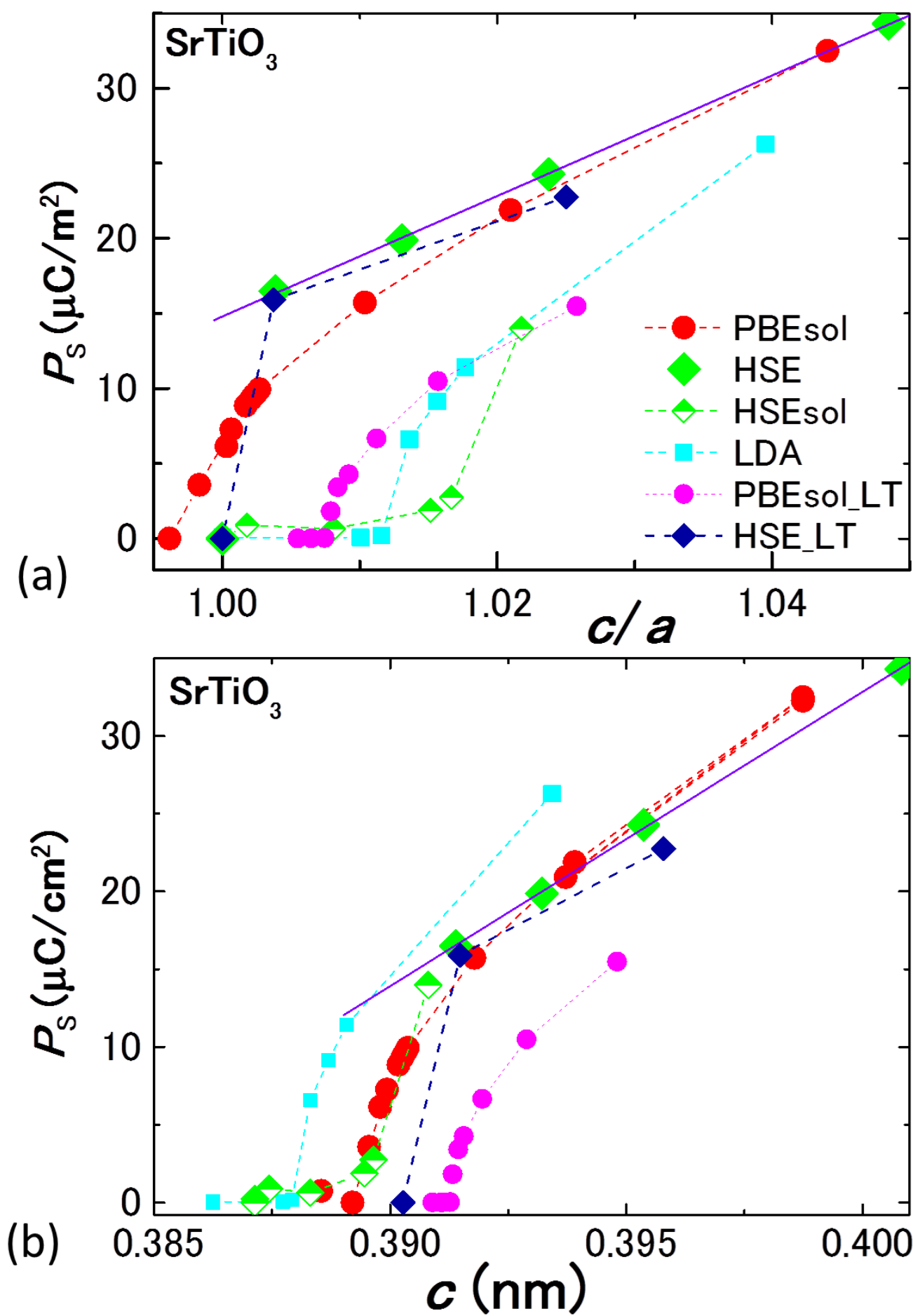

**Fig.5**

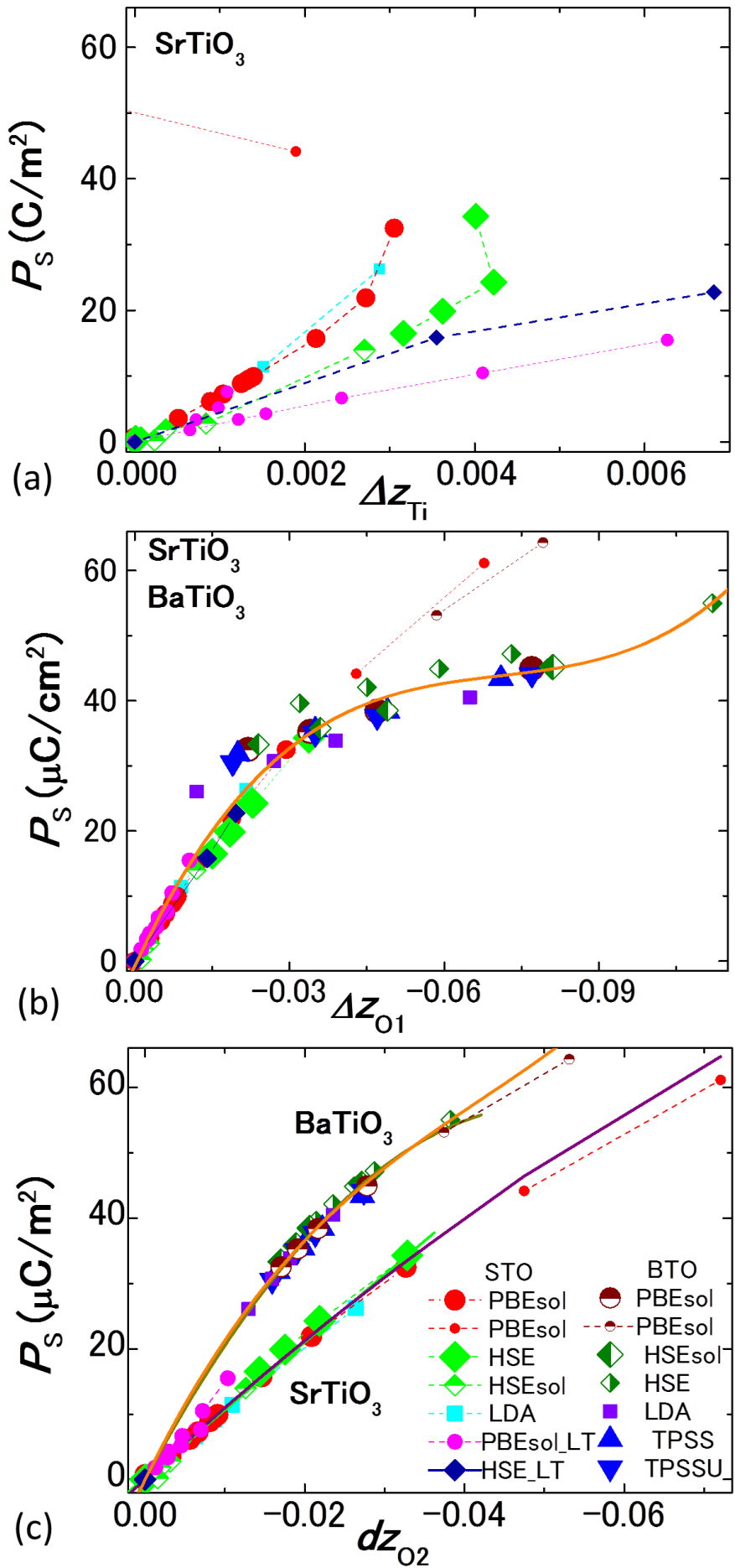

**Fig.6**

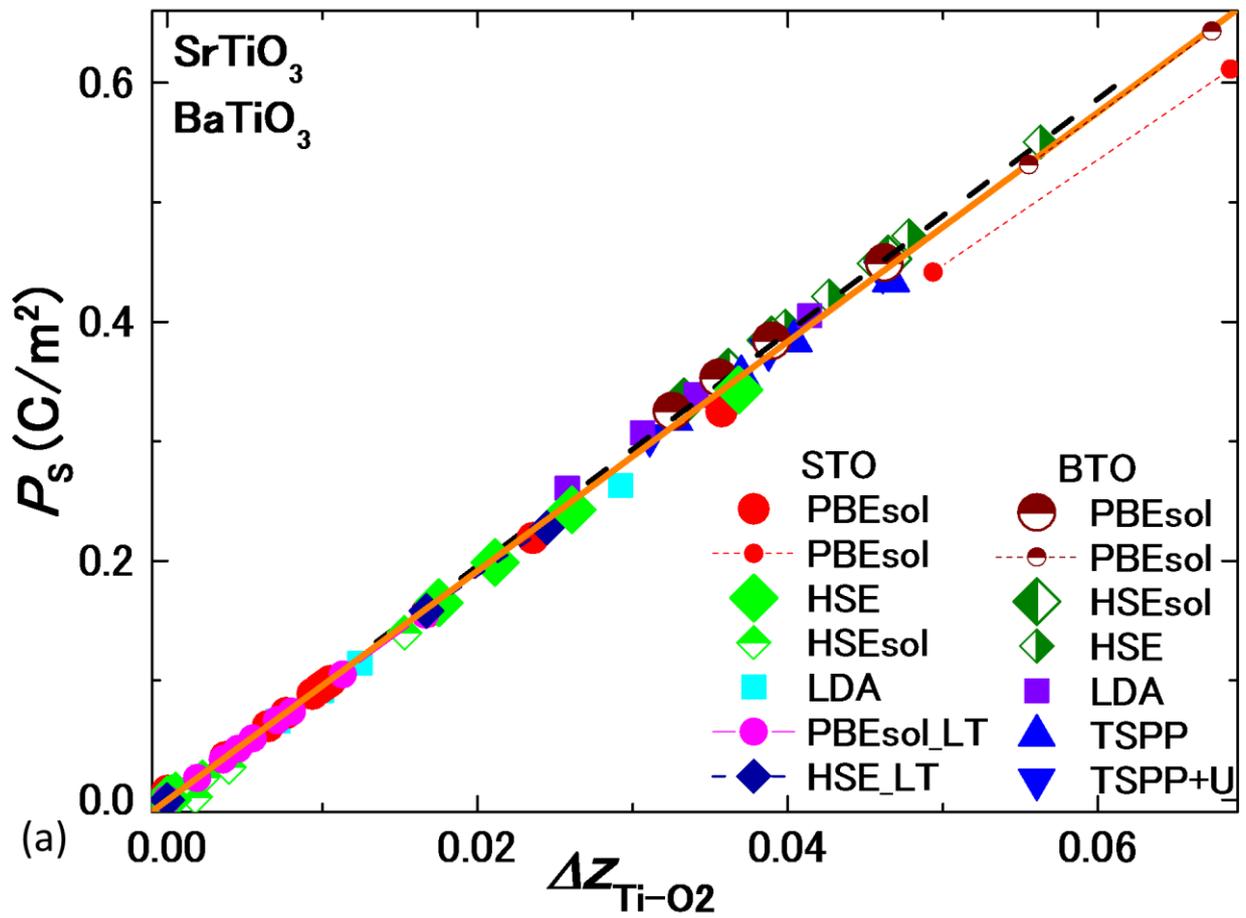

(a)

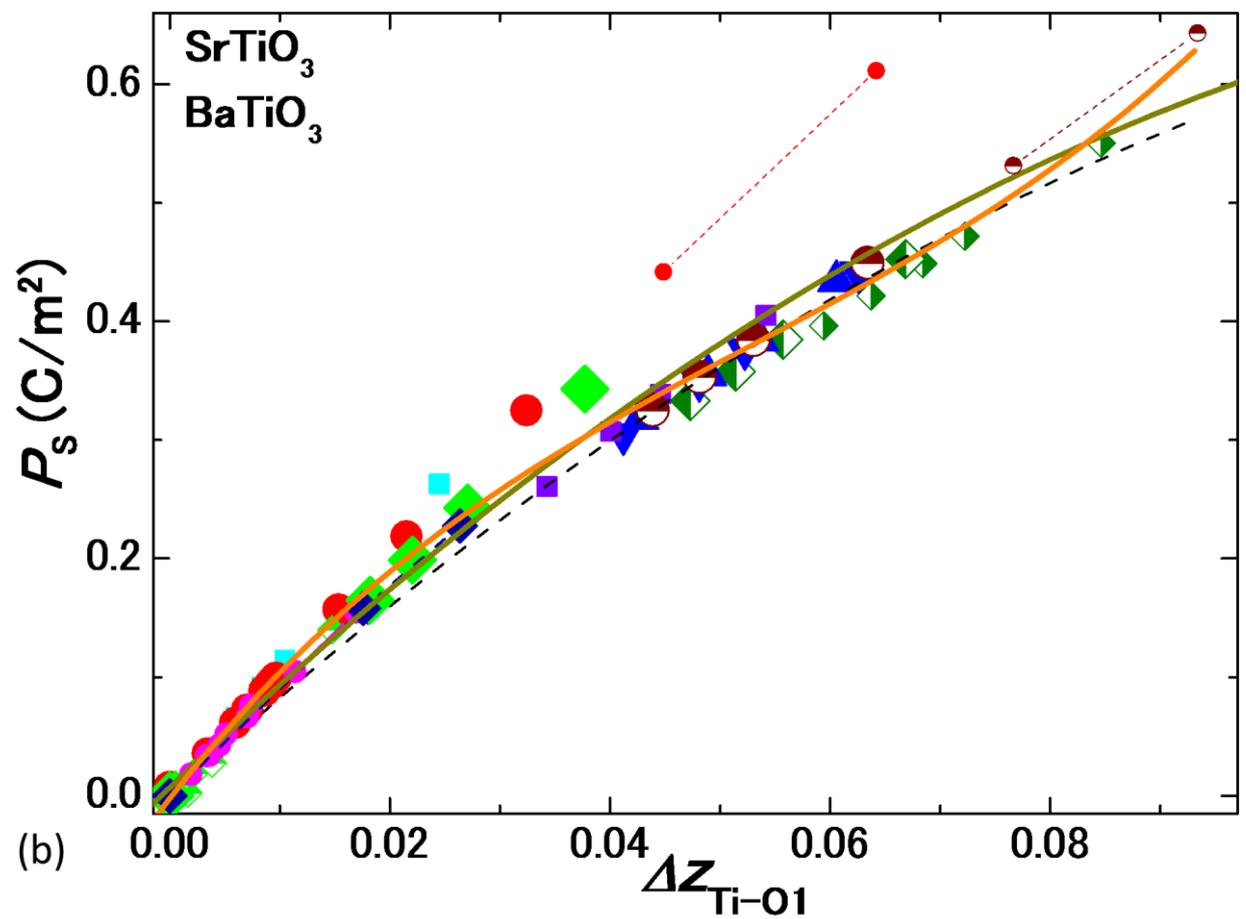

(b)

**Fig.7**

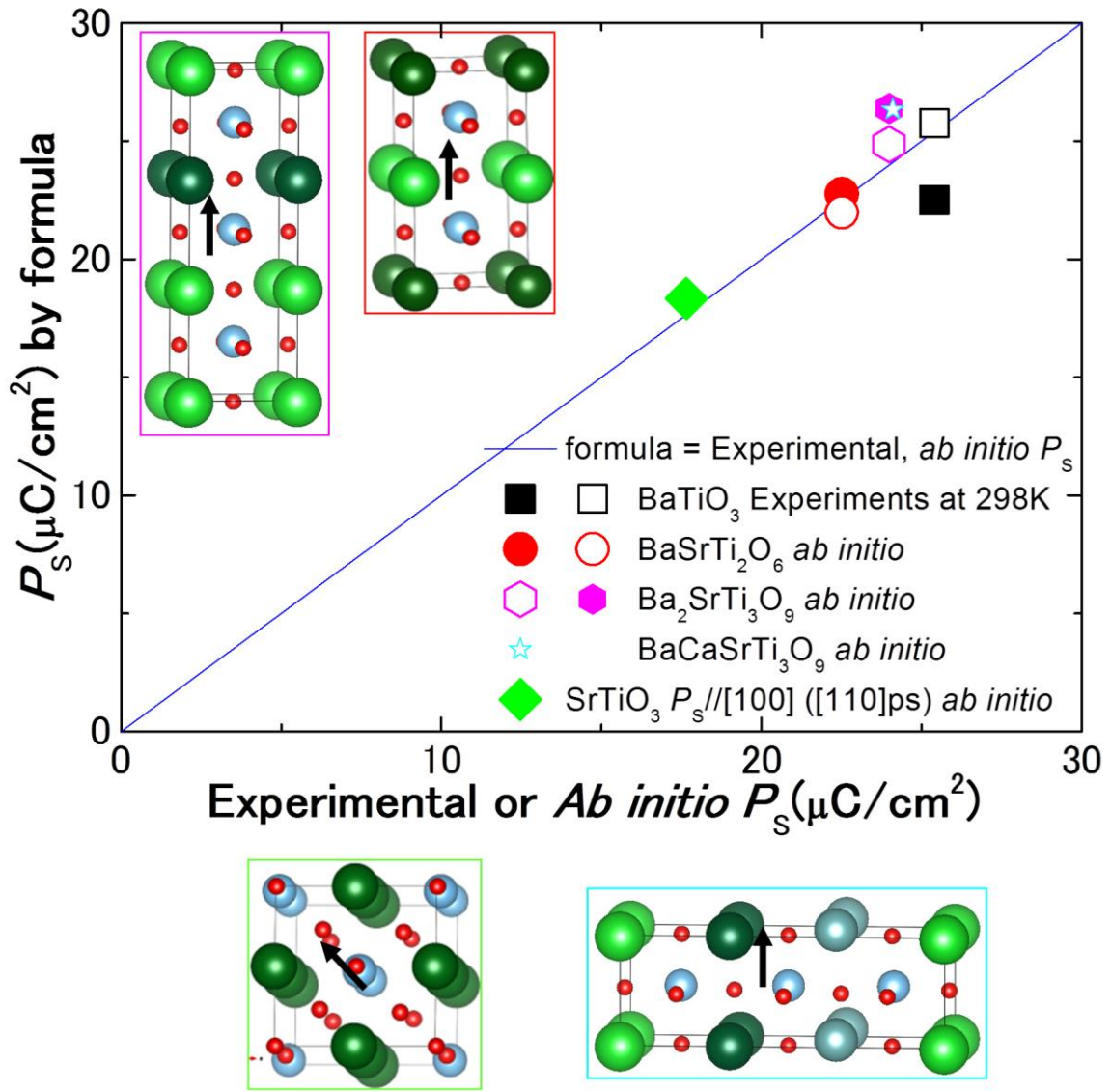

**Fig.8**

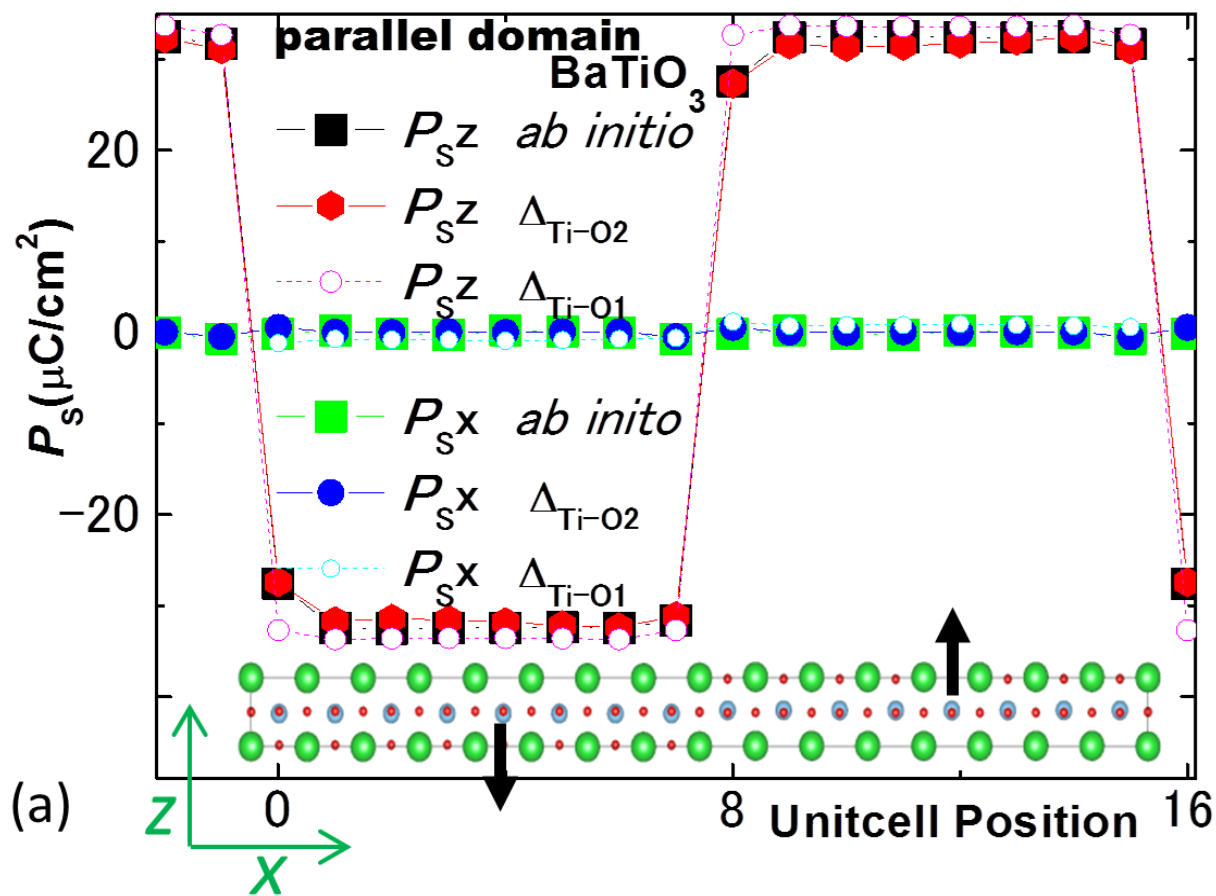

(a)

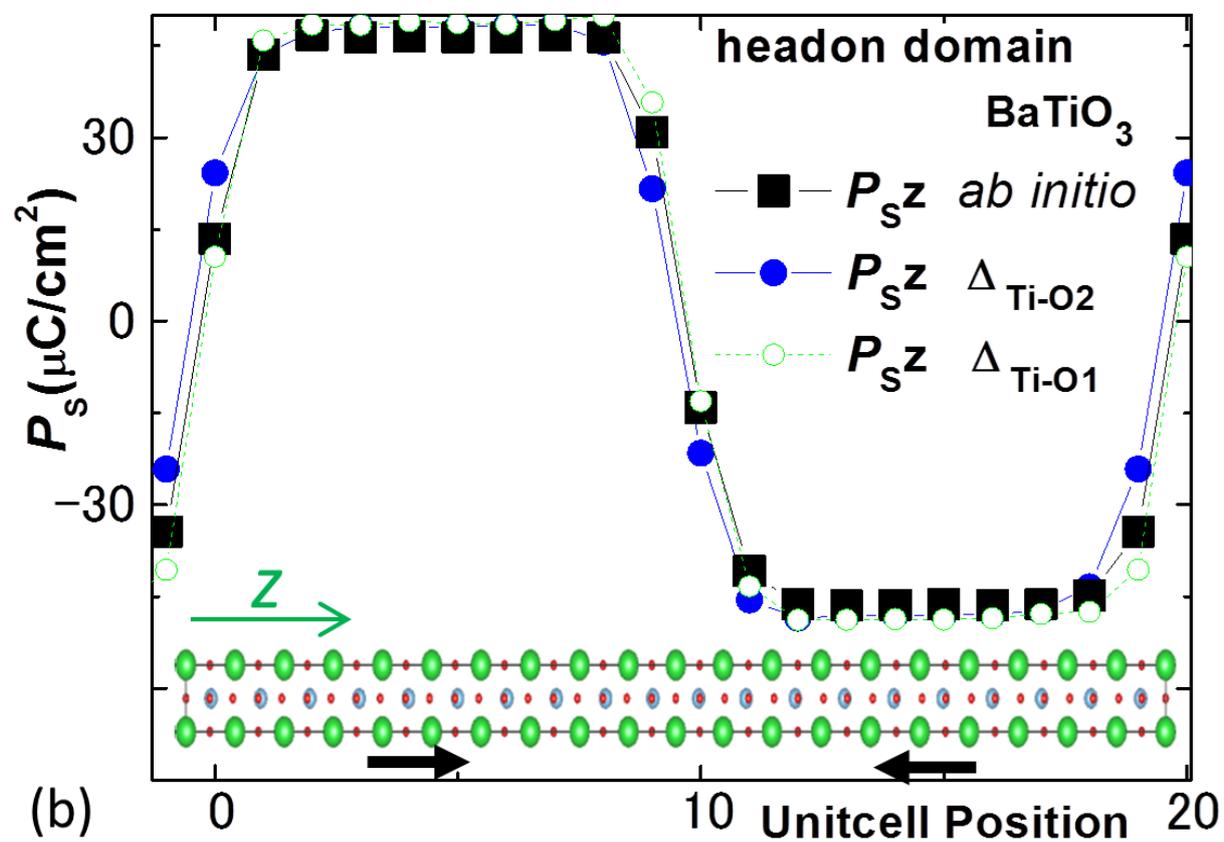

(b)